\documentclass[sigconf]{acmart}
\usepackage{subcaption}
\usepackage{multirow}
\usepackage{booktabs}
\usepackage{pifont}
\usepackage{enumitem}
\usepackage{makecell}
\usepackage{algorithm}
\usepackage{listings}
\usepackage{float}
\usepackage{hyperref}

\AtBeginDocument{%
  }

\copyrightyear{2026}
\acmYear{2026}
\setcopyright{cc}
\setcctype{by}
\acmConference[WSDM '26]{Proceedings of the Nineteenth ACM International Conference on Web Search and Data Mining}{February 22--26, 2026}{Boise, ID, USA}
\acmBooktitle{Proceedings of the Nineteenth ACM International Conference on Web Search and Data Mining (WSDM '26), February 22--26, 2026, Boise, ID, USA}
\acmDOI{10.1145/3773966.3777978}
\acmISBN{979-8-4007-2292-9/2026/02}

\begin{document}


\title{SimDiffRec: Semantic Similarity-Guided Diffusion for Contrastive Sequential Recommendation}



\author{Jinkyeong Choi}
\orcid{0009-0001-5469-0487}
\affiliation{
    \institution{Sejong University}
  \city{Seoul}
  \country{Republic of Korea}
}
\email{jjinchoi@sju.ac.kr}

\author{Yejin Noh}
\orcid{0009-0000-5413-4061}
\affiliation{
    \institution{Sejong University}
    \city{Seoul}
    \country{Republic of Korea}
}
\email{shdpwls0114@sju.ac.kr}

\author{Donghyeon Park}
\orcid{0000-0002-9456-9366}
\authornote{Corresponding author.}
\affiliation{
    \institution{Sejong University}
    \city{Seoul}
    \country{Republic of Korea}
}
\email{parkdh@sejong.ac.kr}


\begin{abstract} 
In sequential recommendation systems, data augmentation and contrastive learning techniques have recently been introduced using diffusion models to achieve robust representation learning. However, most of the existing approaches use random augmentation, which risks damaging the contextual information of the original sequence. Accordingly, we propose \textbf{SimDiffRec}: a Semantic \textbf{Sim}ilarity-Guided \textbf{Diff}usion for Contrastive Sequential \textbf{Rec}ommendation. Our framework leverages the similarity between item embedding vectors to generate semantically consistent noise. Moreover, we utilize high confidence scores in the denoising process to select our augmentation positions. This approach more effectively reflects contextual and structural information compared to augmentation at random positions. From a contrastive learning perspective, the proposed augmentation technique, combined with hard negative sampling, provides more discriminative positive and negative samples, simultaneously improving training efficiency and recommendation performance. Experimental results on five benchmark datasets show that \textbf{SimDiffRec} outperforms the existing baseline models. The code of our framework is available at \url{https://github.com/zingyon/SimDiffRec}.
\end{abstract}
%






\maketitle

\section{Introduction}
Sequential recommendation (SR) systems play a crucial role in predicting the next item a user will interact with based on their past behavior. With the emergence of Transformer-based SR models, it has become possible to effectively capture and model patterns in user behavior~\cite{BERT4Rec, SASRec}. Nevertheless, these powerful models are not immune to the fundamental challenge of data sparsity. Because SR requires learning representations that reflect the sequential dependencies of user interactions, learning robust and generalizable representations from limited data remains a significant hurdle~\cite{S3-rec, CL4SRec}.

\begin{figure}[t]
  \centering
  \includegraphics[width=0.45\textwidth]{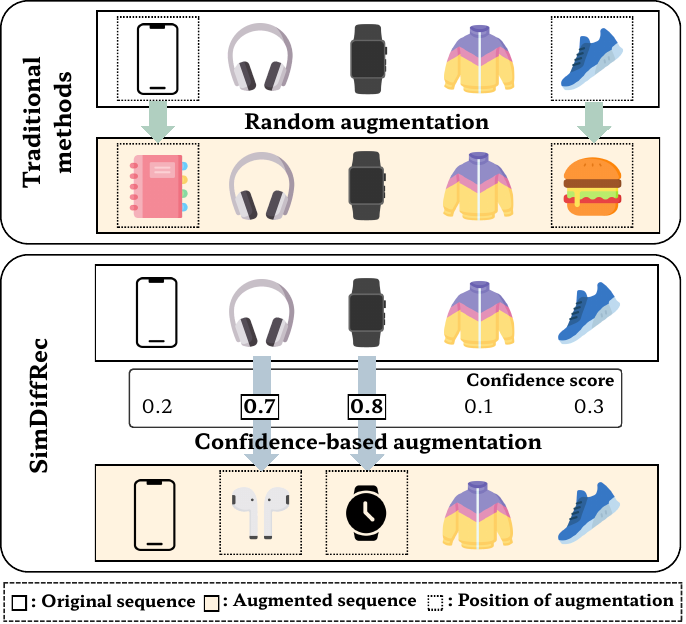}
  \caption{An example comparing traditional random augmentation, which disrupts semantic consistency, with our diffusion confidence-based augmentation that preserves it.} 
\captionsetup{skip=3pt}
\label{fig:fig1}
\end{figure}

To address these issues, data augmentation and contrastive learning (CL) have been actively studied in the SR field~\cite{ICLRec, MCLRec, DuoRec}. However, contrastive learning models involve creating augmented views of a sequence through methods like random item masking, deletion, or noise injection~\cite{EDA, random_aug}. While these techniques can increase data diversity, their reliance on randomness is a critical weakness. As exemplified in the top part of Figure~\ref{fig:fig1}, such random operations often fail to preserve the underlying semantic structure and contextual consistency of the user's behavioral patterns~\cite{CoSeRec, ECL-SR}. Consequently, they risk adding unwanted noisy or even misleading positive samples, leading to information loss and ultimately hindering the model's ability to learn discriminative representations~\cite{DA4Rec}.

Meanwhile, in pursuit of more sophisticated data augmentation, diffusion models~\cite{DDPM, ADM} have recently been utilized in recommendation systems, as well as in CV~\cite{latentdiffusion, photorealistic} and NLP~\cite{Diffusion-lm, DiffusionBERT} domains. Although diffusion models typically use random noise, existing methods that use diffusion for data augmentation have the limitation that the generated samples do not sufficiently maintain semantic consistency or contextual alignment~\cite{context_diff}. To address this problem, the need for more stable and consistent noise instead of random noise has been increasingly highlighted~\cite{random_noise}. This is also true in recommendation systems: failing to adequately reflect the contextual information of the original sequence can lead to degraded contrastive learning performance. In particular, this problem is even more pronounced because it is hard to explicitly model relationships among items and an item’s meaning varies with user intent or context~\cite{cadirec, diffrecsys, DiQDiff}. Therefore, there is a need to introduce structured noise that reflects contextual information instead of randomness.

To achieve this goal, we propose \textbf{SimDiffRec}, a novel data-augmentation framework for Contrastive Sequential Recommendation that leverages a Semantic Similarity–Guided Diffusion model. Crucially, SimDiffRec isn’t a standalone SR model but an augmentation framework that enhances the performance of existing SR models. Instead of relying on randomness, our framework is guided by three core principles to preserve semantic consistency: 
(1) semantic similarity-based noise generation, (2) position selection based on the diffusion model’s reconstruction confidence, and (3) contrastive learning that preserves context and structure.

First, unlike existing approaches that use random noise, SimDiffRec leverages item embeddings from the original sequence to select semantically similar item vectors and uses their average as a structured, meaningful noise. This similarity-based noise generation reduces interference from randomness and enables the production of consistent augmented data without damaging contextual information, even in the early stages of training.

Second, conventional augmentation techniques often select positions randomly, which risks distorting parts of the sequence the model has not yet learned effectively. In contrast, SimDiffRec utilizes the diffusion model's reconstruction confidence to preferentially select positions that have been well-reconstructed. This confidence-based augmentation ensures that augmentation occurs within contexts the model already understands well, preserving both the structural and contextual consistency of the sequence.

By integrating these two strategies with a hard negative sampling technique~\cite{hardneg}, SimDiffRec produces augmented sequences that maximize the effectiveness of contrastive learning. This allows the model to learn more fine-grained differences between the original and augmented sequences, leading to more robust and discriminative representations. When applied to a standard SR model, SimDiffRec demonstrates significant performance improvements on five benchmark datasets, validated its effectiveness and generalizability. 

In summary, the main contributions of this study are as follows:
\begin{itemize}[topsep=5pt, partopsep=0pt, parsep=0pt, itemsep=0pt, leftmargin=1em]
  \item We generate item embedding noise based on semantic similarity instead of random noise, thereby maintaining semantic consistency throughout the diffusion process.
  \item We perform augmentation by focusing on positions where the diffusion model shows high reconstruction confidence, preserving both structural and contextual information.
  \item We combine semantic-based augmentation with hard negative sampling to help the model learn subtle differences and achieve robust discriminative representations.
  \item Through extensive experiments on five benchmark datasets, we verify that the proposed method simultaneously improves recommendation accuracy and generalization performance, and effectively preserves the semantic consistency of sequences.
\end{itemize}

\section{Related Works}
\subsection{Sequential Recommendation}
Sequential recommendation aims to predict items that a user may be interested in the future based on the user’s past interaction sequence. Early sequential recommendation techniques were based on simple probabilistic models such as Markov Chains~\cite{FPMC}. While these methods were effective in modeling short-term dependencies, they had limitations in modeling complex patterns or long-term dependencies in interaction data. These limitations were alleviated with the introduction of RNN-based models capable of processing entire sequences~\cite{GRU4REC2015,GRU4REC2018, HierarchicalRNN}, followed by CNN-based models~\cite{cosrec, caser, NextItNet} that enabled to capture various sequence patterns. Subsequently, attention-based models such as SASRec~\cite{SASRec} applied the Transformer architecture to sequential recommendation, and BERT4Rec~\cite{BERT4Rec} improved performance by modeling bidirectional context. However, they still struggle with data sparsity, which has spurred active research into data augmentation via contrastive learning~\cite{S3-rec,CoSeRec,CL4SRec}.


\subsection{Contrastive Learning Models}
Contrastive learning has emerged as a solution to address the data sparsity problem in SR. It is based on the idea of learning embeddings that place similar items nearby while separating dissimilar ones~\cite{Simcse,PromCSE}. To effectively adopt contrastive learning for sequential recommendation, various methods have been proposed, such as learning item-item relationships~\cite{S3-rec}, data-level augmentation via sequence transformations~\cite{CL4SRec}, and model-level augmentation through generating multiple representations of the input sequence~\cite{DuoRec}. Additionally, contrastive learning has been extended in multiple directions to better accommodate user variability in recommendation tasks. These include the introduction of meta-learning techniques that enable models to adapt to individual user patterns~\cite{MCLRec}, mechanisms for capturing dynamic user intent to handle changing preferences over time~\cite{ICLRec}, and the application of invariance and covariance principles to enhance model robustness against behavioral shifts~\cite{ECL-SR}. However, most existing approaches have limitations in ensuring the semantic consistency of augmented sequences, which can result in unreasonable samples that do not match the context.

\subsection{Diffusion Models} 
Diffusion models have demonstrated strong performance in various fields such as image~\cite{ADM, DDPM, Improved-DDPM} and text generation~\cite{Diffusion-lm, DiffusionBERT, DiffusionCLS, Diffuseq}, and have recently been actively applied to recommendation systems as well. In recommendation systems, diffusion models are utilized in two main ways. The first is as a recommendation model itself, where the diffusion model directly generates the continuous distribution of user preferences and items~\cite{DreamRec, Diffurec, DiffRec, dimerec}. The second approach leverages diffusion models in the data engineering and encoding process to improve recommendation performance, with various methods being proposed. 
In particular, for data augmentation using diffusion models, some methods generate augmented sequences to alleviate data sparsity~\cite{DiffuASR}, combine diffusion models as a plug-in by incorporating the user’s past sequence~\cite{PDRec}, or integrate semantic information into the diffusion process to enhance semantic consistency~\cite{SeeDRec}. Additionally, generating contextually similar sequences in contrastive learning has also been shown to improve the performance of personalized recommendation systems~\cite{cadirec}. However, diffusion-based recommendation models often suffer from contextual distortions in generated sequences due to the use of random noise~\cite{DDRM, DiffRec}. To address this issue, we propose a deterministic diffusion process that preserves contextual consistency in the generated sequences, effectively mitigating such distortions.
\section{Preliminary}
We briefly introduce the sequential recommendation problem and provide an overview of typical sequential recommendation and diffusion models.

\subsection{Problem Statement} 
The goal of sequential recommendation is to predict the item a user will interact with next based on the user’s past interaction sequence. Given a set of users \( U \) and a set of items \( V \), each user \( u \in U \) has a sequence \( S_u = [v_1^u, v_2^u, \dots, v_{|S_u|}^u] \). Formally, the next-item prediction task is formulated as:
\begin{equation}
    \arg\max_{v_i \in V} P(v_{|S_u|+1}^u = v_i \mid S_u)
\end{equation}
where \( P \) denotes the probability of each item being the next in the sequence, and the equation selects the most probable one given \( S_u \).

\subsection{Sequential Recommendation Model}
\label{SR}
Following prior research~\cite{SASRec, cadirec}, we employ a Transformer-based architecture for sequential recommendation. Each item in the set \( V \) is represented by an embedding \( e_k \). Positional embeddings \( p_k \) are added to incorporate order information, resulting in \( h_k^{0} = e_k + p_k \). This sequence is processed through multiple Transformer encoder layers, capturing complex user interaction patterns. The final hidden state \( h_n^{L} \) represents the user's interaction sequence. For prediction, logits for each item \( v_i \) are computed by the inner product between \( h_n^{L} \) and all item embeddings \( M \):
\begin{equation}
r = h_n^{L} M^T
\end{equation}
The binary cross-entropy loss with negative sampling is defined as:
\begin{equation}
\mathcal{L}_{sr} = - \sum_{u\in \mathcal{U}} \sum_{t=1}^n
\left[ \log\big(\sigma(h^L_t \cdot e_{v_{t+1}})\big) + \log\big(1 - \sigma(h^L_t \cdot e_{v_j^-})\big) \right]
\end{equation}
where \( e_{v_{t+1}} \) is the embedding of the ground-truth next item, and \( e_{v_j^-} \) is the embedding of a negatively sampled item. This training approach enhances the model's ability to distinguish between relevant and irrelevant items.

\subsection{Diffusion Models} 
We introduce the general principles of diffusion models based on DDPM~\cite{DDPM}.

\subsubsection{Forward Process}
A continuous embedding sampled from the original data distribution \( \mathbf{x}_0 \sim q(\mathbf{x}_0) \) is gradually noised with standard Gaussian noise over \(T\) steps. This process is modeled as a Markov chain, and the complete forward process is formulated as:
\begin{equation}
\label{eq2}
\begin{aligned}
    q(\mathbf{x}_{1:T}\mid\mathbf{x}_0)&= \prod_{t=1}^{T} q(\mathbf{x}_t\mid\mathbf{x}_{t-1}),\\
    q(\mathbf{x}_t\mid\mathbf{x}_{t-1})&= 
\mathcal{N}(\mathbf{x}_t;\sqrt{1-\beta_t}\,\mathbf{x}_{t-1},\beta_t \mathbf{I})
\end{aligned}
\end{equation}
where \(\beta_t \in (0,1)\) is the variance schedule at time step \(t\), a hyperparameter controlling the noise intensity. After sufficient noise has been added, \( \mathbf{x}_T \) approaches \( \mathcal{N}(0, I) \).

\subsubsection{Reverse Process}
Starting from the final noised embedding \( \mathbf{x}_T \), the original data \( \mathbf{x}_0 \) is gradually denoised using a trained model \( p_\theta \). The reverse process is formulated as:
\begin{equation}
\begin{aligned}
p_{\theta}(\mathbf{x}_{0:T}) &= p(\mathbf{x}_{T}) \prod_{t=1}^{T} 
p_\theta(\mathbf{x}_{t-1}\mid\mathbf{x}_t), \\ 
p_\theta(\mathbf{x}_{t-1}\mid\mathbf{x}_t) &= \mathcal{N} \big(\mathbf{x}_{t-1};
\mu_\theta (\mathbf{x}_t, t), \Sigma_\theta(\mathbf{x}_t, t) \big)
\end{aligned}
\end{equation}
where \(\mu_\theta(\mathbf{x}_t, t)\) and \(\Sigma_\theta(\mathbf{x}_t, t)\) are the predicted mean and variance by the model in the reverse process. The recovered embedding \( \mathbf{x}_0 \) is converted to discrete tokens \( \mathbf{z}_0 \) via rounding or a softmax projection, producing an output sequence.

\section{Methodology}
\label{all_Architecture}
\begin{figure*}[htbp]
    \centering
    \includegraphics[width=0.95\textwidth]{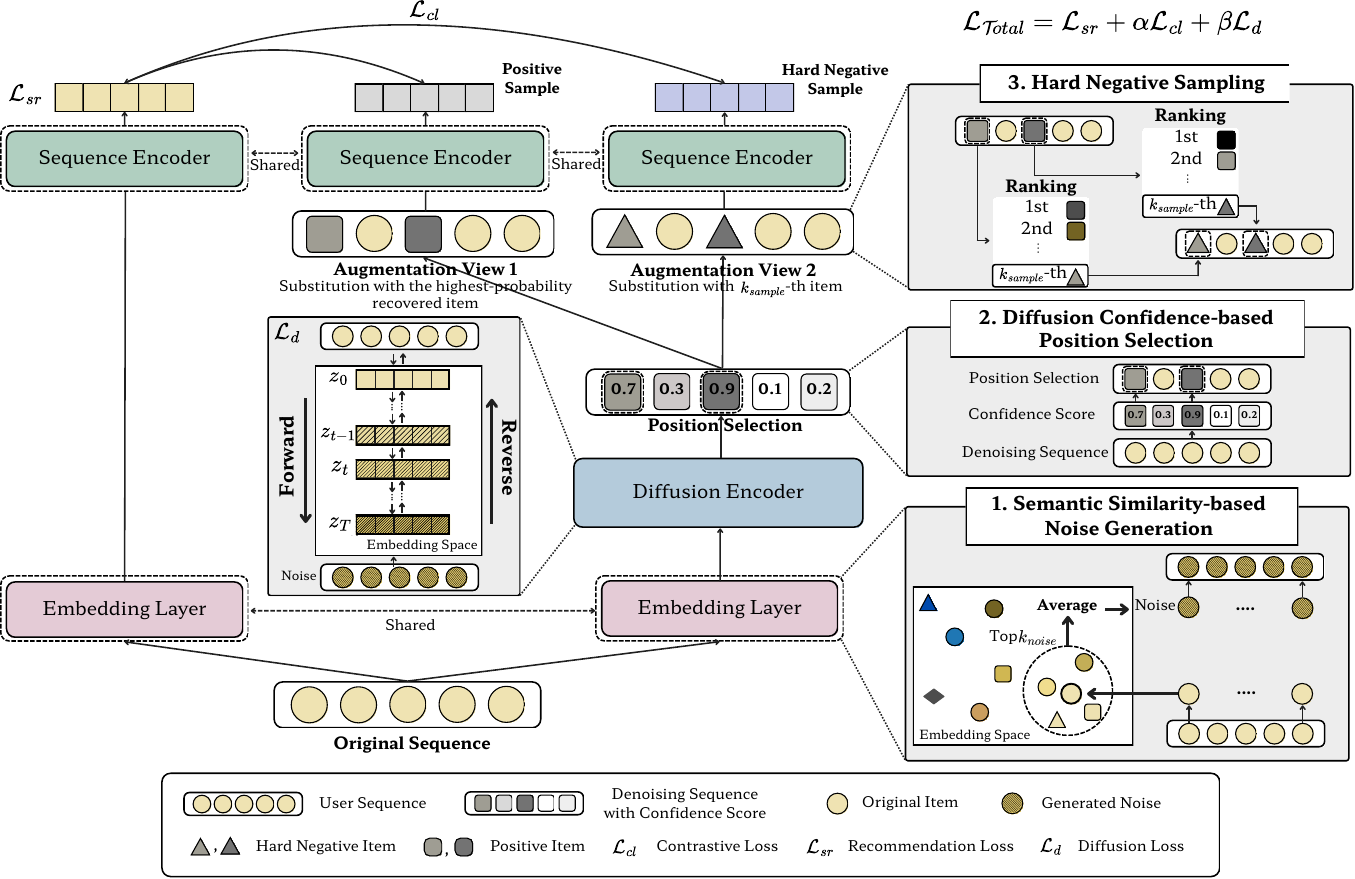}
    \captionsetup{skip=5pt}
    \caption{\small Overview of our proposed SimDiffRec: (1) Embed user–item sequences with a Transformer-based SR model’s embedding layer and insert semantic similarity–based noise. (2) Preserve contextual information during the diffusion denoising process and select augmentation positions with high confidence. (3) Apply hard negative sampling for contrastive learning.}
    \label{fig:all}
\end{figure*}

We propose \textbf{SimDiffRec}, a data augmentation framework designed to enhance the training of SR models, described in Section~\ref{SR}. As illustrated in Figure~\ref{fig:all}, our framework leverages a diffusion model to address contextual information distortion in existing augmentation methods and to perform semantically consistent augmentations. The augmented data are subsequently used for effective CL; we exploit hard negative samples to maximize the SR model’s discriminative capability. Finally, we introduce the end-to-end training objective that unifies these components into a cohesive framework.

\subsection{Semantic Similarity-based Noise Generation}
\label{semantic}
In this study, to maximize augmentation effectiveness while preserving as much of the original sequence’s contextual and semantic information as possible, we propose a noise generation technique that leverages the semantic similarity between item embedding vectors. Existing works perform augmentation by inserting random noise into diffusion models, but this approach can cause excessive distortion of samples in the early stages of training and carries a high risk of losing contextual information.

To address this, instead of highly random noise, we deterministically generate semantically similar noise by utilizing the semantic similarity between item embedding vectors. This method preserves the original semantic information better than random noise and, at the same time, indirectly conveys the original sequence’s contextual information to the diffusion model, thereby minimizing information distortion during augmentation.

Specifically, each item $v_i$ in a user sequence $S_u$ passes through the item embedding layer to obtain an embedding vector sequence $e_u = [e_1, e_2, \dots, e_n]$. Let $W$ be the entire item embedding matrix; we compute a similarity score for each item by taking the dot product between $e_u$ and $W$. Here, to ensure that self-similarity is not included in noise generation, we apply self-masking to exclude an item’s similarity with itself. The similarity is computed as follows:
\begin{equation}
\text{similarity} = e_u \cdot W^T
\end{equation}
Based on the computed similarity scores, items are sorted in descending order, and the top $k_{noise}$ embedding vectors \( e_{j_1}, e_{j_2}, \dots, e_{j_k} \) are selected. The noise is then defined as the average of these vectors:
\begin{equation}
\text{noise} = \frac{1}{k} \sum_{i=1}^{k} \mathbf{e}_{j_i}
\end{equation}

The semantically similar noise generated in this way acts as a guide during the diffusion model’s reconstruction of the sequence, helping prevent it from deviating significantly from the original items. This guidance makes it possible to generate augmentations that are semantically similar to the original items, and reduces the likelihood that the diffusion model will produce samples that overly distort the structure of the original sequence, leading to more effective contrastive learning.

\subsection{Diffusion Forward and Reverse}
We describe the diffusion model procedures for the forward diffusion process, where semantically similar noise is inserted deterministically, and the reverse diffusion process that recovers the sequence.

\subsubsection{Forward Process}
Unlike the standard diffusion model which adds noise sampled randomly from a Gaussian distribution, our approach uses a predefined noise deterministically. Specifically, the Diffusion Encoder takes the original sequence embeddings combined with the semantic noise from Section~\ref{semantic} as its input to start the forward process. At each step $t$, the state $\mathbf{z}_t$ is defined as:
\begin{equation}
\mathbf{z}_t = \alpha_t \mathbf{z}_{t-1} + \beta_t \cdot noise
\end{equation}
where $\alpha_t$ gradually reduces the weight of the original input, and $\beta_t$ controls the strength of the noise. Repeating this process for $T$ steps, the entire forward process is formulated as:
\begin{equation}
\mathbf{z}_T = \prod_{t=1}^{T} \alpha_t \mathbf{z}_0 + \sum_{t=1}^{T} \left( 
\beta_t \cdot noise \prod_{i=t+1}^{T} \alpha_i \right)
\end{equation}

The first term $\prod_{t=1}^{T} \alpha_t \mathbf{z}_0$ represents the original sequence gradually fading, and the second term $\sum_{t=1}^{T} \left( \beta_t \cdot noise \prod_{i=t+1}^{T} \alpha_i \right)$ shows the accumulation of noise added in later steps has a greater impact. This semantic similarity-based noise approach preserves semantic information better than random sampling.

\subsubsection{Reverse Process}
The reverse process proceeds as in the standard diffusion model, by iteratively removing noise. Starting from $\mathbf{z}_T$, the model is formulated with the following conditional probability:
\begin{equation}
p_{\theta}(\mathbf{z}_{t-1} \mid \mathbf{z}_t) = 
\mathcal{N}\Big(\mathbf{z}_{t-1}; \mu_{\theta}(\mathbf{z}_t, t), 
\Sigma_{\theta}(\mathbf{z}_t, t)\Big)
\end{equation}
where $\mu_\theta$ and $\Sigma_\theta$ are learnable functions that the model is trained to output in order to remove noise from the sequence while preserving contextual information.

The entire reverse process is formulated as:
\begin{equation}
  p_{\theta}(\mathbf{z}_{0:T}) = p(\mathbf{z}_T) \prod_{t=1}^{T} 
p_{\theta}(\mathbf{z}_{t-1} \mid \mathbf{z}_t)
\end{equation}
where $p(\mathbf{z}_T)$ is the distribution of the final state with the most noise, and the model uses it as a starting point to reconstruct $\mathbf{z}_0$ as the denoised initial sequence representation.

Finally, we apply the trainable rounding technique~\cite{Diffusion-lm} to map the continuous embedding vector $\mathbf{z}_0$ back to discrete items:
\begin{equation}
p_\theta(s \mid z_0) = \prod_{i=1}^{n} p_\theta(v_i \mid z_i)
\end{equation}
this step helps accurately reconstruct the original item sequence from the continuous representation.

\subsubsection{\texorpdfstring{Loss Function \textbf{L\textsubscript{d}}}{Loss Function L\_d}}
The diffusion model is trained with a loss function designed to minimize the difference between the original sequence and the recovered sequence. Following DiffuSeq~\cite{Diffuseq}, we simplify it as:
\begin{equation}
\mathcal{L}_{\text{d}} = \sum_{t=2}^{T} \|{z}_0 - {f}_\theta(z_t, t)\|^2 + \|{e} - 
{f}_\theta(z_1, 1)\|^2 - \log p_\theta(s \mid z_0)
\end{equation}
Overall, this loss function trains the model by jointly considering the error at each step of the denoising process and the probabilistic discrepancy in the final mapping to the discrete sequence.

\subsection{Diffusion Confidence-based Positioning for Augmentation}
\label{confidence}
Existing data augmentation techniques often incorporate some contextual information in the augmented items, but the position where augmentation is applied is determined randomly. This approach might fail to account for the overall context adequately, and if a poorly learned region is selected, there is a possibility of choosing an incorrect item, leading to contextual distortion. 

To address this problem, we propose a method for selecting augmentation positions based on confidence scores obtained during the diffusion model’s denoising process. Specifically, the confidence score \( c_i \) denotes the probability of the most likely item at position \( i \), computed from the predicted item distribution after denoising the fully noised sequence using the diffusion model. This quantitatively represents how much the model trusts the contextual information at that position. The computation proceeds as follows. First, for each position \(i\), we take the encoder output \(h_i\) and perform a dot product with the item embedding matrix \(W\) to compute a logit vector \(z_i\):
\begin{equation}
z_i = W \cdot h_i
\end{equation}
then, we apply a softmax to obtain the item reconstruction probability distribution \(p_i\) for position \(i\):
\begin{equation}
p_{ij} = \frac{\exp(z_{ij})}{\sum_{k=1}^{|V|} \exp(z_{ik})}, 
\quad j \in V
\end{equation}
from this probability distribution, we take the highest probability value and define it as the confidence \(c_i\) for that position:
\begin{equation}
c_i = \max_{j \in V} p_{ij}
\end{equation}

A high confidence score \(c_i\) indicates that the context at that position is well captured and the model can reconstruct the correct item with high probability. In contrast, a low \(c_i\) implies that the model is less confident about that position, suggesting that the position might not be suitable for augmentation. Using this approach, we compute \(\{c_1, c_2, \dots, c_n\}\) for the entire sequence, then sort them in descending order of confidence and select the top \(k\) positions as the final augmentation targets. By selecting positions based on the predicted confidence, we can reduce the contextual distortion that may occur with conventional random position augmentation, and provide more meaningful augmentations from the perspective of contrastive learning. Consequently, this leads to the generation of more effective positive samples, contributing to improved contrastive learning performance.

\subsection{Hard Negative Sampling for CL}
\label{subsec:hard_negative_cl}
One of the critical factors in contrastive learning is the construction of positive and negative samples. In particular, hard negative sampling plays an important role in providing richer training signals to the model~\cite{hardneg}. In this work, we propose a method to construct hard negative samples in combination with the confidence-based augmentation approach.

First, for a given augmentation position, we select the item reconstructed with the highest probability as the positive sample, as detailed in Section~\ref{confidence}. For that same position, the hard negative sample is then constructed by selecting the item corresponding to the \(k_{sample}\)-th rank from the same probability distribution. Here, \(k_{sample}\) is a rank index;  for instance, if  \(k_{sample} = 1\), the same item as the positive sample is chosen. As \(k_{sample}\) increases, a hard negative sample with relatively high similarity that is difficult to distinguish from the positive sample is selected. This process helps the model learn subtle differences and enhance its discriminative ability.

Using the constructed hard negatives, we optimize an InfoNCE-based contrastive objective~\cite{infoNCE}. Specifically, let \( \mathbf{e}_u \) denote the sequence representation of user \(u\). Let \( \mathbf{e}_{v^+} \) and \( \mathbf{e}_{v^-} \) denote the sequence representations obtained by replacing the items at the designated augmentation positions with the positive and hard-negative items, respectively, where these items are obtained via the diffusion-based augmentation process described above. Let \(\mathcal{N}_b\) denote the set of in-batch negative samples. Then, the contrastive loss is formulated as:
\begin{equation}
\resizebox{0.91\hsize}{!}{$
\mathcal{L}_{cl} = -\log \frac{\exp(\text{sim}(\mathbf{e}_u, \mathbf{e}_{v^+}) / \tau)}
{\exp(\text{sim}(\mathbf{e}_u, \mathbf{e}_{v^+}) / \tau) + 
\exp(\text{sim}(\mathbf{e}_u, \mathbf{e}_{v^-}) / \tau) + \sum\limits_{\tilde{v}
 \in \mathcal{N}_b} \exp(\text{sim}(\mathbf{e}_u, \mathbf{e}_{\tilde{v}}) / 
\tau)}
$
}
\end{equation}
where \(\text{sim}(\mathbf{e}_a, \mathbf{e}_b)\) denotes the cosine similarity between two embedding vectors, and \(\tau\) is a temperature parameter controlling the model’s discrimination sensitivity.

Finally, the hard negative-based contrastive loss \(\mathcal{L}_{\text{cl}}\) is ultimately combined with the sequential recommendation loss \(\mathcal{L}_{\text{sr}}\) and the diffusion loss \(\mathcal{L}_{\text{d}}\) to form the overall training objective:
\begin{equation}
\mathcal{L}_{\text{Total}} = \mathcal{L}_{\text{sr}} + \alpha \mathcal{L}_{\text{cl}} + \beta \mathcal{L}_{\text{d}}    
\end{equation}
where \(\alpha\) and \(\beta\) are hyperparameters that control the contribution of each loss term. By reflecting predicted confidence in augmentation position selection, and using diffusion-based sequences alongside hard negative samples, the approach reduces context distortion from random augmentations and provides semantically meaningful enhancements. This guided to more effective positive sample generation and improves contrastive learning performance.

\section{Experiments}
\subsection{Experimental Settings}
\subsubsection{Datasets}
We evaluated the performance of sequential recommendation using five real-world datasets: Beauty, Toys, Sports, Yelp, and MovieLens (ML-1m). Dataset statistics are shown in Table~\ref{dataset}. The Beauty, Sports, and Toys datasets, collected from Amazon\footnote{http://jmcauley.ucsd.edu/data/amazon/}, contain product purchase and review records. The Yelp\footnote{https://www.yelp.com/dataset} dataset includes user visits and reviews for local businesses, representing implicit feedback. MovieLens\footnote{https://grouplens.org/datasets/movielens/} is a widely used benchmark for evaluating recommendation algorithms based on movie ratings. Following prior work~\cite{cadirec, ECL-SR}, we filtered out users and items with fewer than five interactions to ensure data quality and experimental stability. For the data splitting, we adopt the standard leave-one-out strategy, which is also a common practice in sequential recommendation. For each user's sequence, the last item serves as the test set, the second-to-last as the validation set, and the rest for training.

\subsubsection{Evaluation Metrics}
We evaluate model performance using the following ranking metrics. HR@k (Hit Ratio) measures whether the ground-truth item appears within the top-$k$ recommendations. In other words, it evaluates whether the actual item is among the top-$k$ recommended items. NDCG@k (Normalized Discounted Cumulative Gain) is a ranking quality metric that awards higher scores when the correctly ranked item appears higher in the recommendation list. It reflects not only the presence of the ground-truth item but also its ranking position.

\subsubsection{Baseline Models}
To evaluate the effectiveness of our proposed approach, we compared it against nine baseline models, which can be broadly categorized into Classical Methods, Contrastive Learning-based Methods, and Diffusion-based Methods.
\begin{itemize}[leftmargin=1em]
    \item \textbf{SASRec~\cite{SASRec}}: A Transformer-based SR model using self-attention.
    \item \textbf{BERT4Rec~\cite{BERT4Rec}}: Employs a BERT-style Masked Language Modeling approach to learn contextual user behaviors.
    \item \textbf{CL4SRec~\cite{CL4SRec}}: Uses random augmentation (three strategies) to form positive views for contrastive learning.
    \item \textbf{DuoRec~\cite{DuoRec}}: A dual-level contrastive learning model that enhances performance with model-level augmentation and hard positive sampling.
    \item \textbf{MCLRec~\cite{MCLRec}}: Combines data-level and model-level augmentations to learn multiple user intents.
    \item \textbf{ECL-SR~\cite{ECL-SR}}: Adopts invariance and equivariance based augmentations with a conditional discriminator to learn powerful user sequence representations.
    \item \textbf{DiffuASR~\cite{DiffuASR}}: Predicts the next item by generating pseudo items with a diffusion model, which are then combined with the original sequence to form an extended sequence.
    \item \textbf{DreamRec~\cite{DreamRec}}: Predicts the next item by generating personalized oracle items based on past user interactions.
    \item \textbf{CaDiRec~\cite{cadirec}}: Uses user behavior context to generate recommended items, and provides personalized recommendations tailored to user preferences.
\end{itemize}

\begin{table}[t]
\renewcommand\arraystretch{1.0}
\centering
\caption{The Statistics of Datasets.}  
\setlength\tabcolsep{4.2pt} 
\scalebox{0.9}{
\begin{tabular}{c c c c c c}
\toprule
Datasets & \#Users  & \#Items  & \#Actions & Avg. Length & Sparsity  \\
\midrule
Beauty &22,363 &12,101  &198,502  &8.8  &99.93\%   \\
Toys &19,412 &11,924 &167,597 &8.6 &99.93\%   \\
Sports &35,598&   18,357 & 296,337 &8.3  &99.95\%    \\
Yelp &30,431 & 20,033 &316,354 &10.4 &99.95\%  \\
ML-1m &6,040  &3,953  &1,000,209  &165.6  &95.81\%  \\
\bottomrule
\end{tabular}
}
\label{dataset}
\end{table}

\subsubsection{Implementation Details}
All models were implemented under the same conditions for a fair comparison. In our method, the SR model is evaluated using the RecBole~\cite{recbole} framework and is based on a Transformer architecture with two layers and two attention heads per layer. In contrast, the diffusion model consists of a Transformer with 1 layer and 2 attention heads, and uses a fixed total of 1000 diffusion steps. The hyperparameters for the two loss terms, $\alpha$ and $\beta$, were tested over the range $[0.1, 0.2, 0.3, 0.4, 0.5]$. The dropout rate on the embedding matrix and attention matrix is set to 0.5 for Sports and Toys, and 0.2 for ML-1m and Yelp. The model was trained for 300 epochs with a batch size of 256 and the Adam optimizer at a learning rate of $1e-4$. Following most previous works~\cite{cadirec, ECL-SR}, we set the maximum sequence length to 50 for the three Amazon datasets and Yelp, and to 200 for ML-1M.

\begin{table*}[!htbp]
\renewcommand\arraystretch{0.9}
\centering
\caption{Performance comparison of baseline models on five datasets. Each row’s highest score is highlighted in bold to indicate a statistically significant improvement, while the second-best score is underlined. All experiments were conducted ten times, and each experiment was run until the variance dropped below 0.01.}

\label{tab:methodcomparenew}
\setlength\tabcolsep{3.0pt}
\scalebox{1.0}{
\begin{tabular}{c| l |c c| c c c c| c c c c| c}
\toprule
\multicolumn{2}{c}{ } & \multicolumn{2}{c}{Classical Method} & \multicolumn{4}{c}{Contrastive Learning-based Method} & \multicolumn{4}{c}{Diffusion-based Method} & \multicolumn{1}{c}{ } \\
\cmidrule(lr){3-4} \cmidrule(lr){5-8} \cmidrule(lr){9-12}
Dataset & Metric & BERT4Rec & SASRec & CL4SRec & DuoRec & MCLRec & ECL-SR & DiffuASR & DreamRec & CaDiRec & \textbf{SimDiffRec} & improv. \\
\midrule
\multirow{4}{*}{Beauty} 
  & HR@5  & 0.0178 & 0.0303 & 0.0456 & 0.0471 & 0.0499 & 0.0489 & 0.0445 & 0.0477 & \underline{0.0521} & \textbf{0.0572} & 9.78\% \\
  & HR@10 & 0.0309 & 0.0557 & 0.0774 & 0.0799 & 0.0836 & 0.0822 & 0.0759 & 0.0805 & \underline{0.0867} & \textbf{0.0932} & 7.49\% \\
  & ND@5  & 0.0109 & 0.0163 & 0.0259 & 0.0268 & 0.0286 & 0.0287 & 0.0252 & 0.0273 & \underline{0.0300} & \textbf{0.0310} & 3.33\% \\
  & ND@10 & 0.0152 & 0.0245 & 0.0361 & 0.0374 & 0.0394 & 0.0395 & 0.0353 & 0.0378 & \underline{0.0411} & \textbf{0.0427} & 3.89\% \\
\midrule
\multirow{4}{*}{Toys} 
  & HR@5  & 0.0123 & 0.0304 & 0.0437 & 0.0454 & 0.0475 & 0.0425 & 0.0428 & 0.0456 & \underline{0.0494} & \textbf{0.0560} & 13.36\% \\
  & HR@10 & 0.0202 & 0.0526 & 0.0721 & 0.0748 & 0.0776 & 0.0710 & 0.0707 & 0.0748 & \underline{0.0804} & \textbf{0.0896} & 11.44\% \\
  & ND@5  & 0.0075 & 0.0164 & 0.0236 & 0.0245 & 0.0257 & 0.0241 & 0.0231 & 0.0246 & \underline{0.0267} & \textbf{0.0288} & 7.87\% \\
  & ND@10 & 0.0058 & 0.0236 & 0.0328 & 0.0340 & 0.0354 & 0.0333 & 0.0321 & 0.0341 & \underline{0.0367} & \textbf{0.0396} & 7.90\% \\
\midrule
\multirow{4}{*}{Sports} 
  & HR@5  & 0.0068 & 0.0187 & 0.0272 & 0.0287 & 0.0297 & 0.0283 & 0.0266 & 0.0285 & \underline{0.0309} & \textbf{0.0338} & 9.39\% \\
  & HR@10 & 0.0122 & 0.0343 & 0.0453 & 0.0476 & 0.0484 & 0.0473 & 0.0445 & 0.0469 & \underline{0.0500} & \textbf{0.0562} & 12.40\% \\
  & ND@5  & 0.0041 & 0.0107 & 0.0158 & 0.0166 & 0.0173 & 0.0172 & 0.0154 & 0.0165 & \underline{0.0180} & \textbf{0.0191} & 6.11\% \\
  & ND@10 & 0.0101 & 0.0157 & 0.0221 & 0.0233 & 0.0240 & 0.0233 & 0.0217 & 0.0231 & \underline{0.0249} & \textbf{0.0252} & 1.20\% \\
\midrule
\multirow{4}{*}{Yelp} 
  & HR@5  & 0.0256 & 0.0358 & 0.0414 & 0.0425 & 0.0430 & \underline{0.0438} & 0.0405 & 0.0416 & 0.0431 & \textbf{0.0477} & 8.88\% \\
  & HR@10 & 0.0414 & 0.0508 & 0.0616 & 0.0613 & 0.0647 & \underline{0.0662} & 0.0583 & 0.0601 & 0.0624 & \textbf{0.0689} & 4.08\% \\
  & ND@5  & 0.0164 & 0.0265 & 0.0295 & 0.0306 & 0.0304 & 0.0294 & 0.0293 & 0.0299 & \underline{0.0308} & \textbf{0.0332} & 7.79\% \\
  & ND@10 & 0.0214 & 0.0314 & 0.0353 & 0.0367 & 0.0364 & 0.0366 & 0.0350 & 0.0359 & \underline{0.0370} & \textbf{0.0401} & 8.38\% \\
\midrule
\multirow{4}{*}{ML-1m} 
  & HR@5  & 0.0356 & 0.0770 & 0.1090 & 0.1119 & 0.1181 & 0.1194 & 0.1067 & 0.1136 & \underline{0.1227} & \textbf{0.1278} & 4.16\% \\
  & HR@10 & 0.0533 & 0.1399 & 0.1888 & 0.1936 & 0.2028 & 0.2079 & 0.1853 & 0.1958 & \underline{0.2098} & \textbf{0.2182} & 4.01\% \\
  & ND@5  & 0.0197 & 0.0431 & 0.0620 & 0.0636 & 0.0674 & 0.0675 & 0.0606 & 0.0647 & \underline{0.0701} & \textbf{0.0712} & 1.57\% \\
  & ND@10 & 0.0158 & 0.0625 & 0.0876 & 0.0898 & 0.0947 & 0.0961 & 0.0858 & 0.0911 & \underline{0.0983} & \textbf{0.0996} & 1.32\% \\
\bottomrule
\end{tabular}
}
\end{table*}

\subsection{Overall Comparison}
Table~\ref{tab:methodcomparenew} compares the average performance of our proposed SimDiffRec framework against various existing sequential recommendation methods. The results show that our method consistently achieves higher performance than existing approaches on key metrics like HR@5 and HR@10. To understand the source of this performance gain, we analyze the results by categorizing the baseline models.
\begin{itemize}[leftmargin=1em]
\item \textbf{Classical Methods} (SASRec, BERT4Rec): These methods do not employ contrastive learning or data augmentation, show relatively lower performance compared to contrastive learning-based models. This suggests that contrastive learning enables more effective learning of user sequence representations, thereby improving recommendation performance.
\item \textbf{CL-based Methods} (CL4SRec, DuoRec, MCLRec, ECL-SR): By leveraging contrastive learning to learn the subtle differences between positive and negative samples, they demonstrate improved performance over classical methods. This approach has demonstrated its effectiveness in alleviating the chronic data sparsity problem in the SR domain by securing sufficient positive pairs through data augmentation. However, because they do not sufficiently reflect contextual information during augmentation, they sometimes produce unreasonable positive pairs.
\item \textbf{Diffusion-based Methods}(DiffuASR, DreamRec, CaDiRec): For more sophisticated data augmentation, recent studies have utilized Diffusion models, showing potential for generating more complex and diverse samples. Nevertheless, most of these approaches still fundamentally rely on random noise sampled from a Gaussian distribution. This process can damage the semantic or structural consistency of the original sequence, causing information loss in a context-critical task like recommendation.

\item In contrast, \textbf{SimDiffRec} introduces semantically similar noise by leveraging the similarities between item embeddings and using the average embedding of semantically similar items as noise. Furthermore, by leveraging the diffusion model's denoising and the confidence scores for each position, we select highly confident positive samples and provide a strong learning signal to distinguish subtle differences via hard negative sampling. As a result, this approach not only preserves the contextual consistency of the original sequence but also reliably outperforms existing methods across multiple benchmarks.

\end{itemize}

\begin{table}[b]
\centering
\caption{Ablation study on three datasets.}
\setlength\tabcolsep{4.0pt}
\scalebox{0.85}{
\begin{tabular}{c|c|cccc}
\toprule
 Dataset & Metric & {\makecell[c]{w/o $k_{noise}$}} & {\makecell[c]{w/o $c_{aug}$}} & {\makecell[c]{w/o $k_{sample}$}} & Ours \\
\midrule
\multirow{4}{*}{Beauty}  
  & HR@5    & 0.0558 & 0.0539 & 0.0551 & 0.0572 \\
  & NDCG@5  & 0.0306 & 0.0295 & 0.0300 & 0.0310 \\
  & HR@10   & 0.0911 & 0.0900 & 0.0921 & 0.0932 \\
  & NDCG@10 & 0.0420 & 0.0412 & 0.0419 & 0.0427 \\
\midrule
\multirow{4}{*}{Toys}  
  & HR@5    & 0.0555 & 0.0537 & 0.0560 & 0.0560 \\
  & NDCG@5  & 0.0283 & 0.0273 & 0.0279 & 0.0288 \\
  & HR@10   & 0.0895 & 0.0882 & 0.0876 & 0.0896 \\
  & NDCG@10 & 0.0395 & 0.0388 & 0.0385 & 0.0396 \\
\midrule
\multirow{4}{*}{Sports}  
  & HR@5    & 0.0328 & 0.0328 & 0.0332 & 0.0338 \\
  & NDCG@5  & 0.0169 & 0.0172 & 0.0171 & 0.0191 \\
  & HR@10   & 0.0523 & 0.0544 & 0.0559 & 0.0562 \\
  & NDCG@10 & 0.0232 & 0.0242 & 0.0244 & 0.0252 \\
\bottomrule
\end{tabular}
}
\label{tab:ablation}
\end{table}

\subsection{Ablation Study}
We conducted an ablation study to evaluate the impact of each component on the overall model performance, with the results presented in Table~\ref{tab:ablation}.
First, "\textit{w/o} \(k_{noise}\)" removes our semantic similarity-based noise. This removal led to decreases in HR@10 and NDCG@10, indicating that leveraging item embedding similarities to preserve the structural and semantic characteristics of the original sequence plays an important role in effective data augmentation.
Next, "\textit{w/o} \(c_{aug}\)" removes the confidence-based positioning strategy, which relies on softmax probabilities, and instead selects augmentation positions randomly. This led to a consistent performance drop across all datasets, indicating that the augmentation position selection itself is a critical factor governing performance. This module contributes to learning stability by guiding augmentation toward contexts the model already confidently understands, thereby inducing the generation of reliable positive samples.

Finally, "\textit{w/o} \(k_{sample}\)" is the setting where hard negative sampling is excluded from the contrastive learning process. In this case, a distinct performance degradation was observed, particularly in NDCG. This demonstrates that negative samples that are difficult to distinguish play a decisive role in enhancing the model's discriminative ability. Using hard negatives provides a much stronger contrastive learning signal.
Consequently, each component individually contributes to the performance improvement of the proposed framework, and removing any one of them degrades performance. These results empirically substantiate that our three core modules achieve a synergistic effect.

\subsection{Hyperparameter Study}
\captionsetup{skip=0pt}
\begin{figure}[t]
    \centering
    \begin{subfigure}[b]{0.47\linewidth} 
        \centering
        \includegraphics[width=\linewidth, height=4cm]{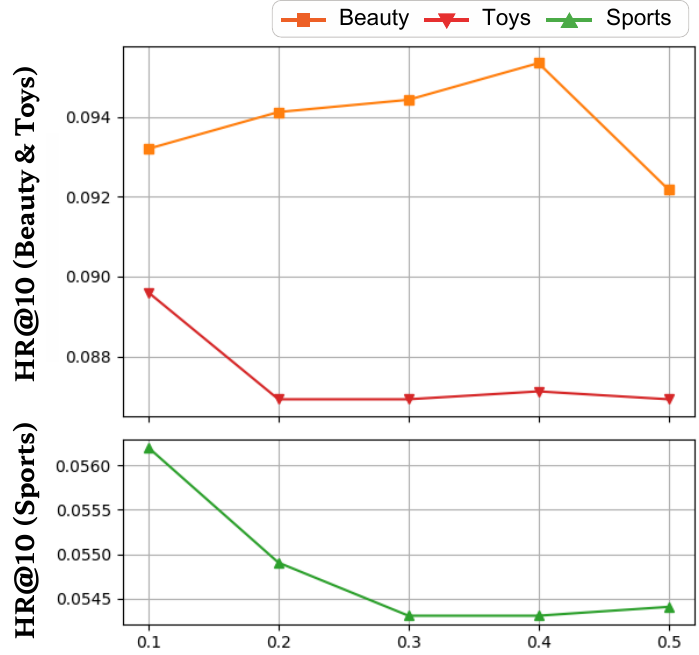}
        \subcaption{Contrastive learning loss $\alpha$} 
    \end{subfigure}
    \begin{subfigure}[b]{0.47\linewidth}
        \centering
        \includegraphics[width=\linewidth, height=4cm]{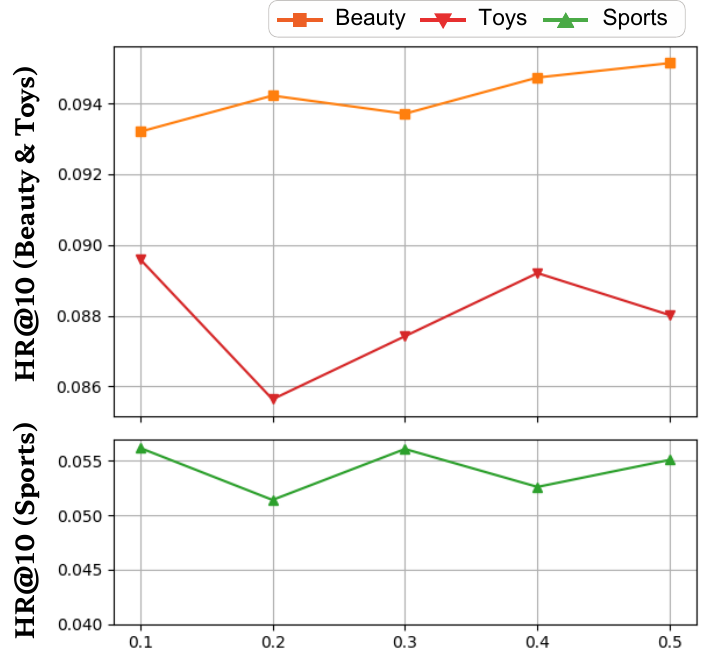}
        \subcaption{Diffusion loss $\beta$} 
    \end{subfigure}
    \vspace{5mm}
    \caption{Hyperparameter study of $\alpha$, $\beta$ on three datasets.}
    \vspace{-5mm}
    \label{fig:figure3}
\end{figure}

We analyze the influence of two key hyperparameters on HR@10: the contrastive loss weight (\(\alpha\)) and the diffusion loss weight (\(\beta\)) as shown in Figure~\ref{fig:figure3}. As crucial hyperparameters, their values are fixed before training and tuned on a validation set. They determine the relative contribution of each loss term to the overall objective function, making the search for an optimal balance essential for maximizing model performance. For \(\alpha\), in most datasets performance gradually improved up to \(\alpha = 0.1\) and reached an optimum, but when \(\alpha\) exceeded 0.2, performance tended to decline. In the Beauty dataset in particular, the best performance was achieved at \(\alpha = 0.4\), suggesting that differences in sparsity compared to other datasets influenced the effectiveness of contrastive learning. By contrast, \(\beta\) did not significantly affect HR@10 in most cases, although the Beauty dataset showed a temporary performance dip around \(\beta = 0.2\). Nevertheless, overall performance remained stable. In summary, the Beauty dataset is optimized at relatively higher loss weights around (\(\alpha, \beta = 0.4, 0.4\)), whereas the Toys and Sports datasets achieve the best performance at lower weights (\(\alpha, \beta = 0.1, 0.1\)). This demonstrates that finely tuning the balance between loss terms based on dataset characteristics is essential for maximizing model performance.


\subsection{Analysis of Hard Negative Sampling and Similarity Embeddings}
\begin{figure}[t]
    \centering
    \begin{subfigure}[t]{\linewidth} 
        \centering
        \includegraphics[width=\linewidth]{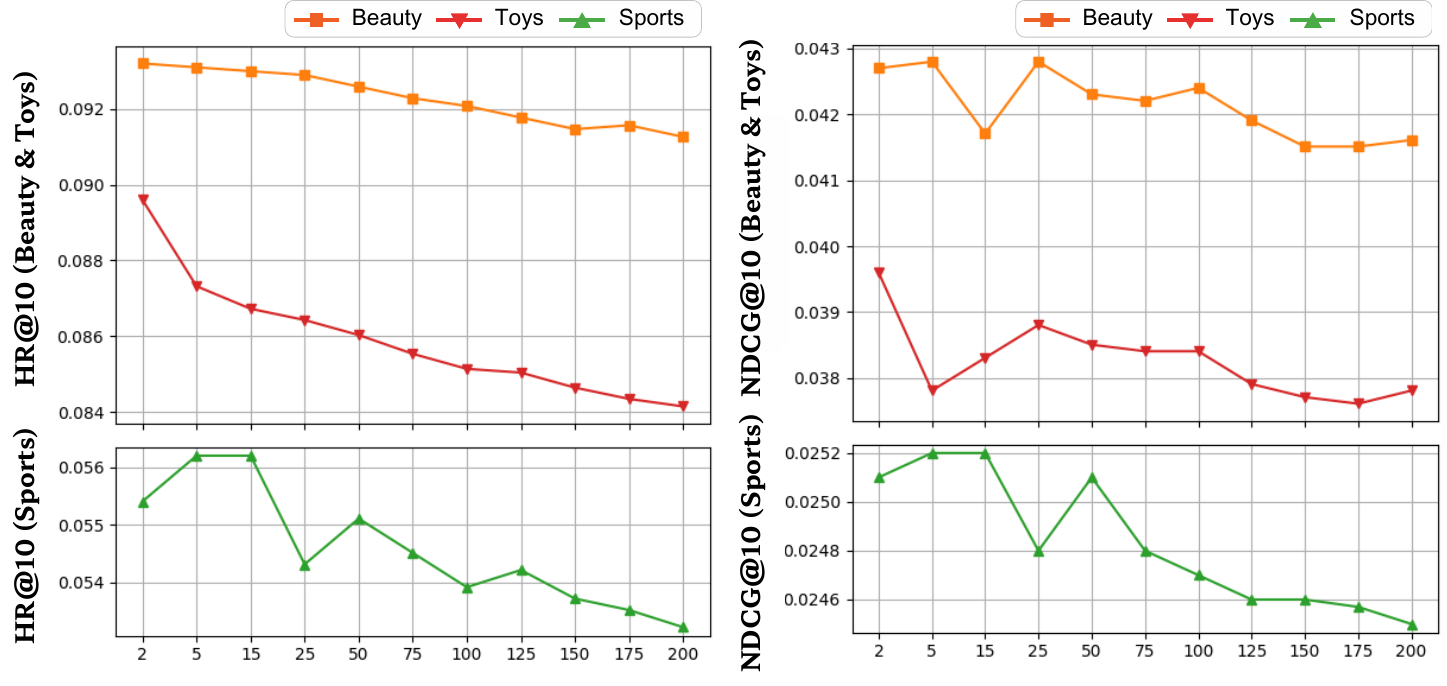}
        \subcaption{Number of hard negative rank}
    \vspace{5mm}
    \end{subfigure}
    \begin{subfigure}[t]{\linewidth}
        \centering
        \includegraphics[width=\linewidth]{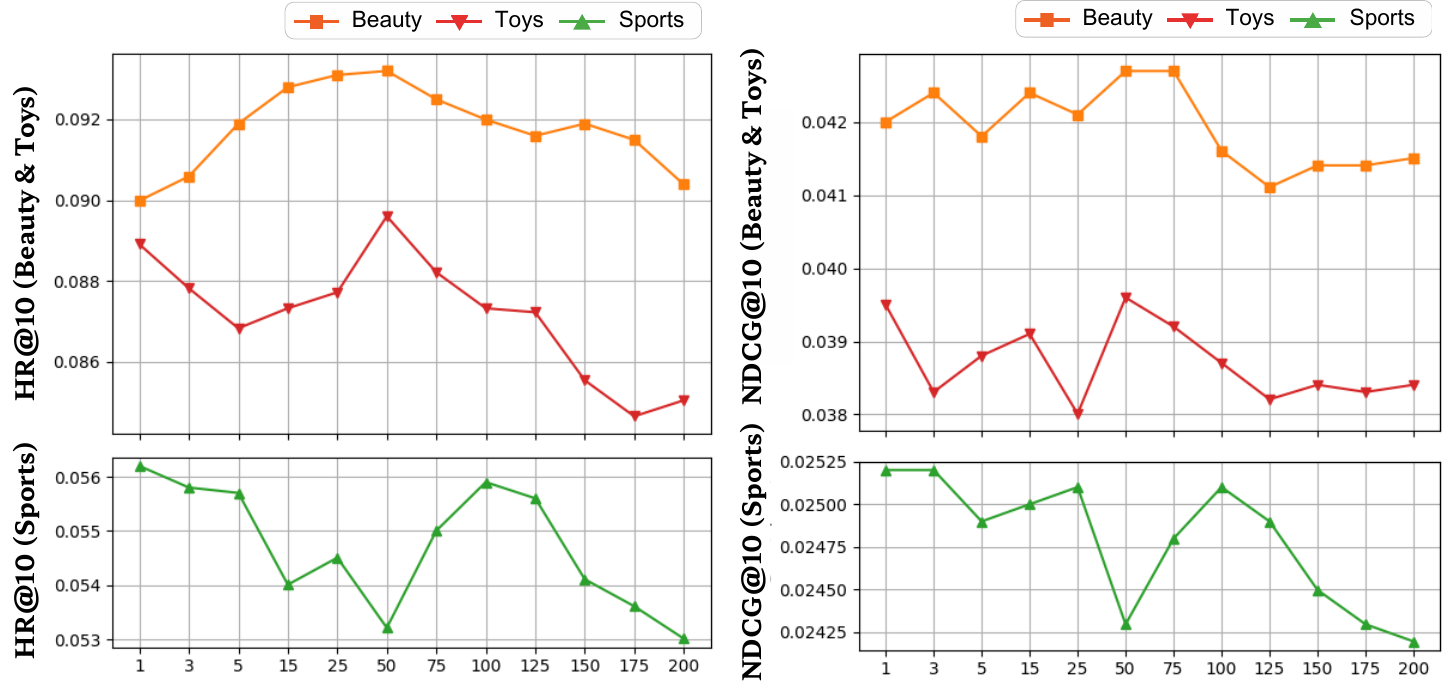}
        \subcaption{Number of similar embeddings} 
    \end{subfigure}
    \vspace{5mm}
    \caption{Performance changes based on hard negative sampling rank and similarity-based noise.}
    \label{fig:figure4}
\end{figure}

We conducted two experiments to assess the model’s robustness. First, we adjusted the ranking index (\(k_{sample}\)) for hard negative sampling to analyze how the model is affected by the similarity between positive samples and negative samples. Next, we varied the number of similar embeddings (\(k_{noise}\)) to evaluate the impact of semantic similarity–based noise on recommendation performance. These experiments were performed on the Beauty, Toys, and Sports domains, and the results are presented in Figure~\ref{fig:figure4}.

As \(k_{sample}\) increases, generating negative samples that are further from the positive samples, overall performance decreased in all domains. In contrast, when \(k_{sample}\) is smaller, the negative samples become more similar to the positive samples, allowing the model to learn subtle differences that are harder to distinguish, which in turn tends to improve performance. However, the extent of this effect varied depending on domain characteristics, and when \(k_{sample}\) was set too high, the role of hard negatives was diminished. Furthermore, the analysis of \(k_{noise}\) revealed that in the low item sparsity domains (Beauty and Toys), optimal performance was achieved when \(k_{noise}\) ranged from 25 to 50, whereas in the high sparsity Sports domain, the best performance was observed with \(k_{noise}\) between 1 and 5. These findings suggest that properly balancing the composition of negative samples and similar items in contrastive learning-based recommendation systems can effectively mitigate the challenges posed by data sparsity and noise.

\setlength{\textfloatsep}{10pt}

\begin{figure}[t]
\centering 
\includegraphics[width=0.9\linewidth]{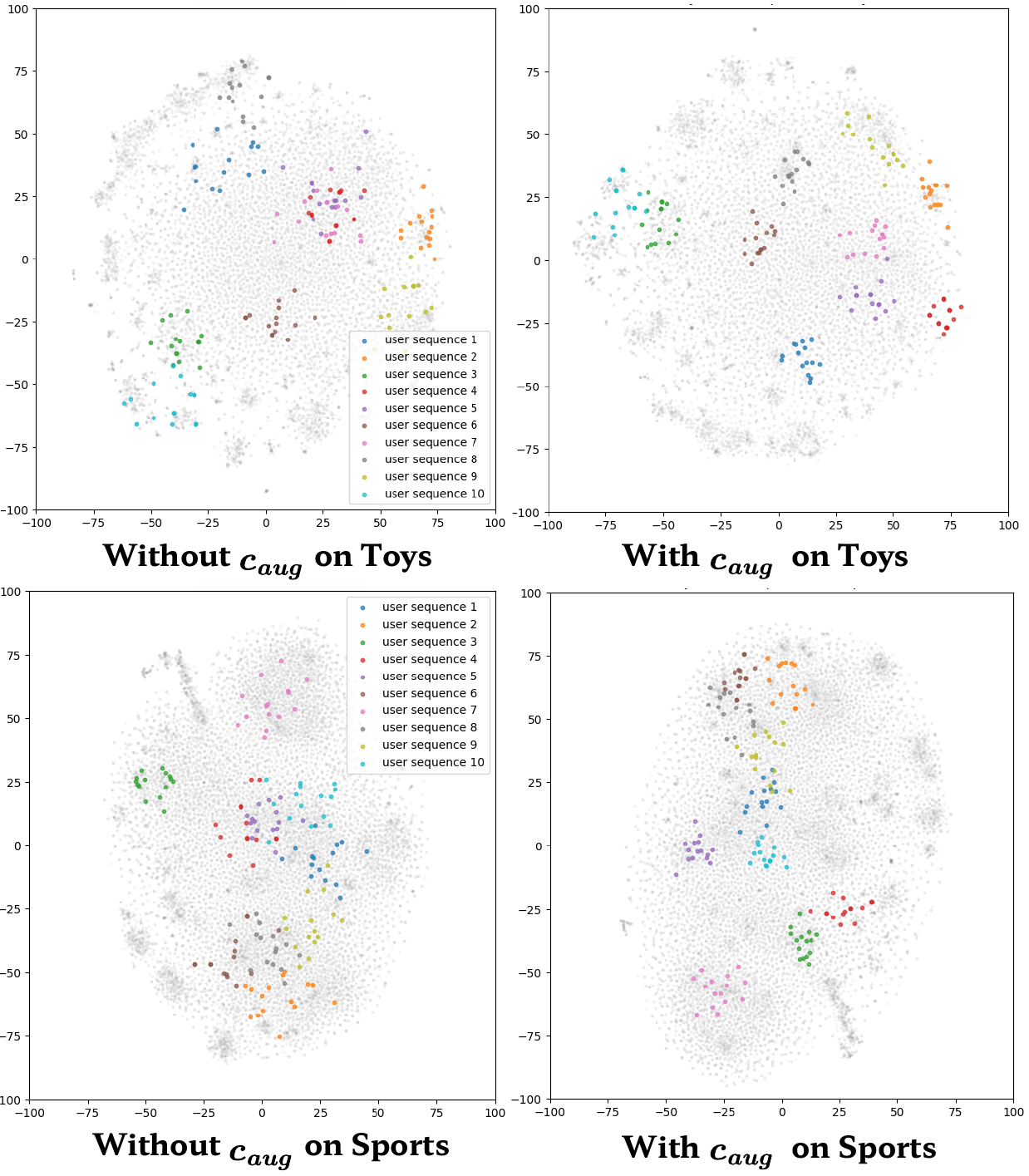}
\vspace{5mm}
\caption{T-SNE visualization of sequence embeddings trained without and with $\boldsymbol{c}_{\boldsymbol{aug}}$.}
    \label{fig:figure5}
\end{figure}

\subsection{Sequence Representation Visualization}
To analyze the impact of confidence-based augmentation ($c_{aug}$) on item representations within user sequences, we visualized item embeddings using T-SNE~\cite{TSNE} for cases with and without confidence-based augmentation, as shown in Figure~\ref{fig:figure5}. The conventional random position augmentation does not incorporate any contextual information into the augmentation position, making it more likely to be applied to relatively under-trained parts. This can lead to the degradation of contextual information or distortion of the characteristics of the user behavior sequence. In fact, when random augmentation was applied, the similarity between items within a sequence was low, and clusters tended to be somewhat dispersed. In contrast, confidence-based augmentation performs the augmentation at the position where the model is most confident after diffusion restoration, thereby preserving contextual information while applying meaningful transformations. As a result, the distances between items within the same sequence become closer, item representations are formed more similarly, and they can be effectively utilized for contrastive learning. These findings suggest that confidence-based augmentation contributes to enhancing the consistency of item representation learning by reflecting contextual information more effectively than the conventional random position augmentation.

\subsection{Computational Complexity}
In this analysis, for a fair comparison across all models, we applied each model’s specific techniques but excluded simple augmentation that only increases the amount of data without altering the actual interaction patterns. As shown in Table~\ref{tab:train-infer}, during training, SimDiffRec incurs a similar level of additional cost to other recent augmentation-based models such as ECL-SR~\cite{ECL-SR} and CaDiRec~\cite{cadirec}. However, at inference time, since the diffusion module is not used, it is faster than DreamRec~\cite{DreamRec}, which requires additional sampling using diffusion. As a result, the inference speed of the proposed model is virtually identical to that of the backbone model, SASRec~\cite{SASRec}, with no additional overhead, making it efficient enough for immediate deployment in real-world service environments. All experiments were conducted on a single NVIDIA A100-PCIE 40GB GPU.



\begin{table}[t]
\centering
\small
\setlength{\tabcolsep}{6pt}
\renewcommand{\arraystretch}{1.12}
\caption{Comparison of time costs for training and inference
per epoch (Beauty, Sports, Toys).}
\label{tab:train-infer}
\vspace{2mm}
\begin{tabular}{l*{3}{cc}}
\toprule
\multirow{2}{*}{Methods} &
\multicolumn{2}{c}{Beauty} &
\multicolumn{2}{c}{Sports} &
\multicolumn{2}{c}{Toys} \\
\cmidrule(lr){2-3}\cmidrule(lr){4-5}\cmidrule(lr){6-7}
& Train & Infer & Train & Infer & Train & Infer \\
\midrule
BERT4Rec & 4.82s & 1.92s & 6.74s & 4.10s & 4.56s & 2.22s \\
SASRec   & 4.94s & 0.44s & 6.55s & 1.05s & 4.52s & 0.60s \\
ECL-SR   & 9.67s & 0.71s & 13.73s & 1.98s & 8.92s & 1.08s \\
DreamRec & 9.14s & 0.88s & 13.81s & 6.64s & 9.36s & 3.79s \\
CaDiRec  & 9.03s & 0.45s & 13.70s & 1.06s & 7.02s & 0.66s \\
SimDiffRec     & 9.14s & 0.44s & 13.51s & 1.11s & 7.38s & 0.63s \\
\bottomrule
\end{tabular}
\end{table}

\section{Conclusion}
In this study, we propose \textbf{SimDiffRec}, which leverages semantic similarity between item embeddings for data augmentation instead of random noise. It strategically selects augmentation positions using confidence scores derived from a diffusion model and incorporates hard negative sampling to capture subtle differences in negative samples. Experimental results show that our method outperforms all baseline models, validating the effectiveness of the proposed semantic similarity-based noise generation and confidence-based augmentation strategies.


\section*{Ethical Considerations}
To the best of our knowledge, this research poses no risk of serious harm. SimDiffRec was evaluated on publicly available datasets and does not contain sensitive or personally identifiable information. While the model learns user behavioral patterns and context in depth, its purpose is limited to improving the performance of the general recommendation task of predicting the next item a user will interact with.

\begin{acks}
This research was supported by the Culture, Sports and Tourism R\&D Program through the \grantsponsor{GS_KOCCA}{Korea Creative Content Agency (KOCCA)}{https://www.kocca.kr} grant funded by the Ministry of Culture, Sports and Tourism (MCST) in 2025 (Project Name: Cultivating masters and doctoral experts to lead digital-tech tourism, Project Number: \grantnum{GS_KOCCA}{RS-2024-00442006}) and the Regional Innovation System \& Education (RISE) through the \grantsponsor{GS_RISE}{Seoul RISE Center}{https://www.seoulrise.or.kr}, funded by the Ministry of Education (MOE) and the Seoul Metropolitan Government (Project Number: \grantnum{GS_RISE}{2025-RISE-01-019-04}).
\end{acks}



\bibliographystyle{ACM-Reference-Format}
\balance  
\bibliography{7_references}



\end{document}